\documentclass[conference]{IEEEtran}
\usepackage{amsmath,amssymb,amsfonts,amsthm} 
\usepackage{graphicx,float} 
\usepackage{hyperref}
\usepackage{multirow}
\usepackage{tablefootnote}
\usepackage{flushend}

\usepackage{enumitem, booktabs, textcomp, xcolor, pifont, comment, layout}

\usepackage{caption}
\usepackage{subcaption}

\newcommand{\todo}[1]{\textcolor{red}{TODO: #1}\PackageWarning{TODO:}{#1!}}

\newcommand{\mao}[1]{\textcolor{orange}{#1}\PackageWarning{}{#1!}}

\begin{document}
\title{End-to-End Network Slicing with Traffic Awareness Resource Allocation}


\title{INA-Infra: An Open and Extensible Infrastructure for Intent-driven Network Automation Research}

\author{
\IEEEauthorblockN{Nguyen-Bao-Long Tran, Tuan V. Ngo, Mao V. Ngo, Binbin Chen, Jihong Park and Tony Q. S. Quek}
\IEEEauthorblockA{Singapore University of Technology and Design (SUTD), 487372, Singapore\\ long\_tran@mymail.sutd.edu.sg, \{vantuan\_ngo, vanmao\_ngo, binbin\_chen, jihong\_park, tonyquek\}@sutd.edu.sg
}
\and

}

\maketitle

\begin{abstract}
As telecommunications systems progress to support diverse use cases with heterogeneous and dynamic Quality of Service (QoS) requirements, it becomes an increasingly complex task to automatically manage various resources involved --- from radio, compute, to X-haul network, which are distributed from the edge to the cloud. Intent-driven network automation can play an important role in NextG networks to meet this need. Towards this, we have developed INA-Infra, an open, extensible, and end-to-end 5G/beyond 5G network infrastructure with intent-driven network automation and end-to-end network slicing capability. INA-Infra is designed using open-source components and is based on O-RAN architecture. INA-Infra manages the network infrastructure, various resources, and (virtualized / containerized) network functions using Nephio --- a cloud-native intent automation solution. It also incorporates intent-driven intelligent control using a Resource Management rApp and a Network Slicing xApp. We demonstrate that INA-Infra can manage the 5G network in a highly automatic and optimized manner, allowing the mobile network operators to focus on specifying the intents of different traffic classes.
\end{abstract}


\section{Introduction}
Fifth-generation networks (5G) are transforming our society, by enabling an increasingly diverse set of use cases, from enabling people to interact via extended reality (XR), to connecting smart things. Different use cases can have heterogeneous Quality of Service (QoS) requirements, and these requirements can change rapidly over time in 5G and next-generation (NextG) networks. Network slicing is a key technology in 5G and NextG networks to support such heterogeneous and dynamic use cases. Network slicing enables the creation of multiple virtual networks on a shared physical infrastructure, each tailored to specific application needs. By leveraging network slicing, service providers can dynamically allocate resources to different network slices, ensuring that each slice meets the specific performance requirements of various use cases such as enhanced mobile broadband (eMBB), massive machine type communication (mMTC), and ultra-reliable low latency communication (URLLC) \cite{E2E_Slicing}. It also becomes important to automatically manage the various resources involved in serving these use cases. Not only do the radio resources need to be carefully scheduled, but the compute resources (e.g., to deploy the virtualized / containerized network functions), and the X-haul network resources, need to be coordinated as well. Intent-driven network automation is a promising direction to manage such complexity, where the operators can focus on specifying the intents of different traffic classes, while the network management system can automatically translate to suitable network configurations. This can be a promising direction for NextG research. To support such research, it will require end-to-end and extensible research infrastructures, which can simulate real-world conditions and provide a robust environment for experimentation. These testbeds must support large-scale, cost-effective experimentation while ensuring high throughput, low latency, and reliability across different network slices.


A NextG network consists of the Core Network (CN) and Radio Access Network (RAN) components. At the RAN side, the O-RAN effort represents a paradigm shift from a traditional single-vendor closed system into an open ecosystem, where virtualization of network functions (NFs) together with a cloud-native approach of containerized microservices is used to deliver fine-grained control and management. Combining O-RAN architecture with a cloud-native NextG core network, an end-to-end (E2E) NextG network can offer a large degree of configurability. For example, a cloud-native 5G Core Network can be deployed in a cloud environment with abundant resources, while a 5G Radio Access Network (RAN) can be deployed at the network’s edge to enable real-time, low-latency applications. Open-source communities play a crucial role in developing software components that can be used for NextG research testbeds~\cite{ICTexpress}. Key communities include the O-RAN Software Community (OSC), Open Air Interface (OAI)~\cite{oai_web} Software Alliance (OSA), and the Open Networking Foundation (ONF). 
While these open-source projects provide valuable solutions for NextG research, there is no readily available E2E framework that allows researchers to test and innovate on intent-driven network automation. 
To address this gap, we introduce INA-Infra---Intent-driven Network Automation Infrastructure---an open, extensible, and E2E NextG research framework based on Nephio~\cite{nephio_wiki}, a Kubernetes-based intent-driven automation framework. 
Leveraging open-source projects, INA-Infra (i) realizes high-level intents from mobile network operators (MNOs) into Nephio's languages to manage the distributed network infrastructure from the Edge to the Cloud, and (ii) is based on O-RAN architecture with containerized NFs for different components of RAN, in particular, by using the O-RAN defined RAN Intelligent Controller (RIC) to support intent-driven intelligent control in the RAN. 

The remaining of this paper will present:
\begin{itemize}
    \item Design and implementation of INA-Infa --- an Intent-driven Network Automation Infrastructure, which builds upon multiple open-source software, in particular, Nephio-managed containerized NFs of 5G Core and RAN, and is deployed over a distributed environment. 
    \item Demonstration of using INA-Infra to realize E2E Network Slicing~\cite{ORANWG1_SlicingArch_2024} solution (from Core and RAN container's placement to radio allocation), which uses our Resource Management-rApp and Network Slicing-xApp for managing RAN functions and slicing resources for different use cases.
\end{itemize}



\section{Related Works}


Regarding E2E testbed research, Colosseum \cite{colosseum}, an O-RAN digital twin, can advance algorithms and testing while enabling RAN optimization through its emulation capabilities and support for software stacks. OPEN6GNET \cite{fokus} offers a cloud-native Kubernetes-based solution for deploying E2E 5G networks, facilitating hands-on experimentation and a deeper understanding of 5G architecture for students and researchers. 

Intent-driven network automation is a software-defined approach that translates the intent of operators into network configuration that can achieve such high-level intents. It needs to combine network intelligence, analytics, and orchestration, so that the network operators would not need to code and execute individual tasks manually. Network slicing is a key mechanism for achieving this, while the Nephio project~\cite{nephio_wiki} under the Linux Foundation is an important open-source initiative to achieve cloud native network automation through an intent-driven approach.
For E2E network slicing, ORCA \cite{ORCA2024} has demonstrated a closed-loop slice assurance method using cloud-native Nephio for RAN and Core orchestration. Further fine-grained resource management could be achieved by utilizing the RIC with AI-driven rApps \cite{AIrApps}. 





At the RAN side, some key open-source projects are:
\begin{itemize}
\item OAI~\cite{oai_web} provides core networks (5G CN, EPC CN), 5G RAN, and RIC~\cite{ICTexpress, AIrApps} and SMO (OAI’s MOSAIC5G) for intelligent control and orchestration frameworks.
\item ONF provides SD-RAN ($\mu$ONOS as a Near-Real-Time RIC), a RAN simulator (CU and DU), and an SDK for xApp development~\cite{ICTexpress}.
\item OSC provides the implementation of O-RAN-related components (O-DU Low, O-DU High, Near-Real-Time and Non-Real-Time RIC).
\end{itemize}




\section{INA-Infra Network Slicing Architecture}
\begin{figure}[!t]
\centering
\includegraphics[width=\columnwidth,trim=0 0 0 0.1in]{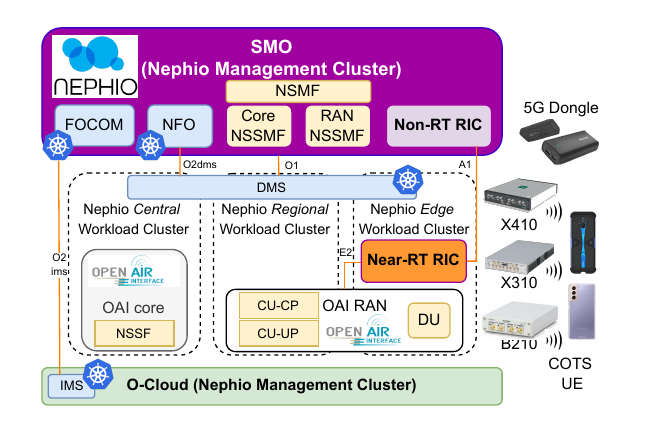} \caption{INA-infra open 5G network testbed system architecture}
\label{fig:nephio-archi}
\vspace{-5mm}
\end{figure}

INA-infra enables MNOs to define network slices SLA from the Service Management Orchestration (SMO) and Nephio. This high-level intention will be translated to low-level action of E2E network slicing through 5G CN and the RAN across distributed infrastructure. 

\subsection{INA-infra Open 5G Network Testbed}
Our INA-infra open 5G Network testbed (Fig. \ref{fig:nephio-archi}) at Future Communication and Connectivities Lab (FCCLab) in Singapore University of Technology and Design (SUTD) is based on open-source software (OAI), and software-defined radio (SDR) platform (USRP B210/X310/X410). 


\begin{figure*}[h]
\centering
\includegraphics[width=0.99\textwidth]{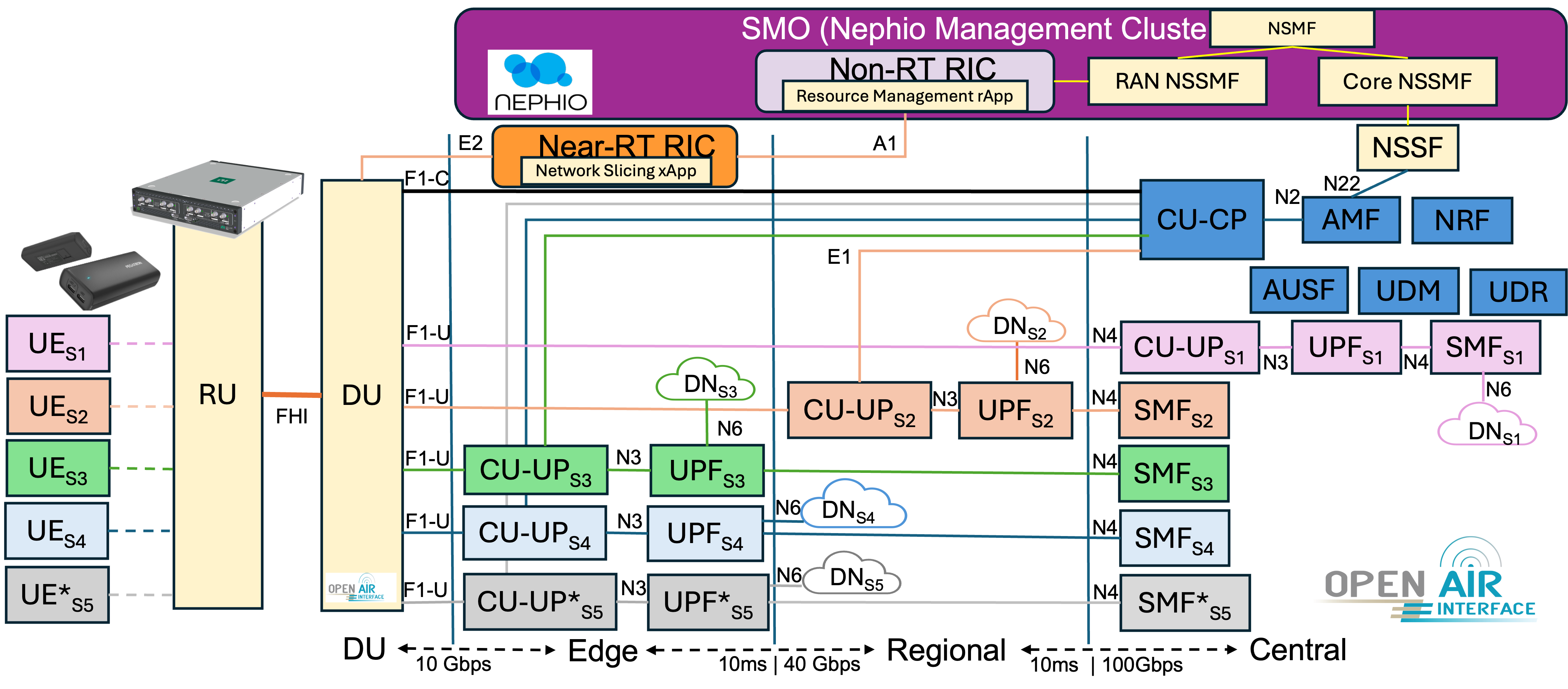} \caption{Experimental setup of multiple network slices across distributed infrastructure.}
\label{fig:Testbed-setup}
\vspace{-3mm}

\end{figure*}

\subsubsection{SMO, O-Cloud \& Non-Real-Time (Non-RT) RIC}
The SMO integrates various management services, allows for the creation of E2E network slices~\cite{ORANWG1_SlicingArch_2024} and the automated management of the entire network. 

The O-Cloud \cite{ORANWG6_O2} manages physical resources such as servers, networks, and storage, and hosts O-RAN functions. It interfaces with the SMO through the O2 interface to oversee infrastructure management and the life cycles of O-RAN network functions, offering two main services: Infrastructure Management Services (IMS), which handles physical resource allocation, inventory maintenance, and communication with the SMO’s Federated O-Cloud Orchestration and Management (FOCOM), and Deployment Management Services (DMS), which works with the Network Function Orchestrator (NFO) to manage network function provisioning, FCAPS, and software lifecycle.

The Non-Real-Time (Non-RT) RIC \cite{ICTexpress} provides two key management and orchestration services: Intent-based Network Management, which enables operators to set network management objectives and policies using a high-level language like rApp, and Intelligence Orchestration, which ensures the proper deployment and conflict-free operation of rApps and xApps.

\subsubsection{Nephio}
Nephio \cite{nephio_wiki} is a Kubernetes-based project aimed at enabling intent-driven automation for network functions and infrastructure. Nephio allows users to express high-level intent, and provides intelligent, declarative automation that can set up the cloud and edge infrastructure, render initial configurations for the network functions, and then deliver those configurations to the right clusters to get the network up and running. Nephio manages three key areas:

\begin{itemize}
    \item Cloud Infrastructure Automation: Utilizes Kubernetes to manage multi-vendor, multi-site networks, simplifying cloud-native infrastructure operations.
    \item Workload Resource Automation: Automates resource allocation and scaling for network workloads, leveraging Kubernetes' extensibility.
    \item Workload Configuration: Uses open-source tools like kpt and ConfigSync for declarative configuration management, ensuring consistency across deployments.
\end{itemize}

\subsubsection{5G Core Network (CN)}
In 5G CN, our environment supports both open-source 5G CN (e.g., OAI 5GC, Open5GS, Free5GC), and commercial 5G CN (e.g., QCT 5GC, Ataya 5GC) that support network slicing features at the core level. OAI 5G CN is used in our experiment.

\subsubsection{Radio Access Network (RAN)}
The RAN consists of Central Unit-Control Plane (CU-CP as a logical node hosting RRC and control part of PDCP), Central Unit-User Plane (CU-UP as a logical node hosting user plane part of PDCP and SDAP), Distributed Unit (DU as a logical node hosting RLC, MAC, High-PHY based on O-RAN split), and Radio Unit (RU as a logical node hosting Low-PHY and RF processing). OAI RAN is used in our experiment.


\subsubsection{Near-Real-Time (Near-RT) RIC}
Leveraging the E2 interface and xApps, the Near-RT RIC~\cite{ICTexpress} enables RAN control, management and optimization within a 10-millisecond to 1-second window. OAI implements FlexRIC~\cite{ICTexpress} as a multi-platform compatible NearRT-RIC. 



\subsection{End-to-End Network Slicing}
Network slicing features allow MNOs to (i) provide services based on customer requirements over a common infrastructure, or (ii) flexible deploy sharing RAN equipment among operators.
A network slice is a logical network with a set of RAN resources (CU-UP, CU-CP, DU, physical radio resource block - PRBs) and core services (including UPF and SMF) over a shared, distributed infrastructure.  A sample network slicing RAN and Core is shown in Fig.~\ref{fig:Testbed-setup}, with some dedicated NFs at RAN or 5G CN (e.g., CU-UP, UPF, SMF) and some NF shared among the network slices (e.g., AMF, CU-CP, DU). 

3GPP System Architecture~\cite{3GPP23501} with Network Slice Selection Function (NSSF) in 5G core supports: selecting Network Slice instances (NSI) serving the UE; determining the allowed NSSAI, and O-RAN working group 1 (WG1) also defines architecture to support network slicing in an open RAN environment~\cite{ORANWG1_SlicingArch_2024}. 

To support E2E network slicing in our testbed, we integrate NSSF in 5G CN, a Resource Management rApp (RM-rApp) in Non-RT RIC~\cite{ICTexpress} in SMO, and a Network Slicing xApp (NS-xApp) in the Near-RT RIC (FlexRIC) to adaptively place and adjust resources for DU, CU-UP, CU-CP, UPF, SMF deployment in different geographic Data Centre (DC) pool (e.g., Central, Regional, Edge). With virtualization and containerized technologies, DU, CU-UP, CU-CP, UPF, SMF are deployed as Kubernetes resources managed by Nephio. Deploying NFs at the Edge DC reduces end-to-end delay but incurs higher costs, while deploying at the Central DC is more affordable but increases end-to-end delay.

3GPP TS 28.531~\cite{3GPP28531} defined management aspects of network slicing such as Network Slice Instance (NSI) which refers to end-to-end network slice, and the Network Slice Management Function (NSMF) which splits the Network Slice Instance (NSI) requirements into individual Network Slice Subnet Instances (NSSI) such as RAN NSSI or CN NSSI. The provisioning of network slicing follows a lifecycle consisting of preparation, commissioning, operation, and decommissioning, during which NSI operations such as creation, activation, modification, deactivation, and termination are performed. 

\subsubsection{CN Slicing Selection}
In 5G CN, the Core Network Slice Subnet Management Function (Core-NSSMF) oversees the management and orchestration of Network Slice Subnet Instances (NSSIs)~\cite{E2E_Slicing}. The Network Slice Selection Function (NSSF) selects the NSI, determines the allowed network slice selection assistance information (NSSAI), and the corresponding AMF to serve the UE. Upon initial UE access, a set of network slices, including an AMF, is assigned based on the requested NSSAI. If the RAN cannot select an AMF, the request is forwarded to a default AMF, which retrieves the UE’s subscription data and works with the NSSF to assign a new AMF. For PDU session establishment, the AMF selects slice-specific functions like the SMF via the NRF, using the S-NSSAI, DNN, and local policies.


\subsubsection{RAN--E2 Nodes}
When an NSSI is requested, the RAN Network Slice Subnet Management Function (RAN-NSSMF) instantiates virtual network functions for CU-UP, CU-CP, and DU to support the slice \cite{E2E_Slicing}. These processes enable slice-aware resource management within the RAN. Additionally, performance measurement reports are provided through the O1 and E2 interfaces.

\subsubsection{Resource Management rApp (RM-rApp)}
The RM-rApp, hosted in the Non-RT RIC, retrieves RAN slice SLA targets from SMO and NSSMF, monitors RAN slice subnet performance via the O1 interface per S-NSSAI, and reacts when SLAs are violated. In such cases, it calculates the optimal compute resources for virtual network functions (like CU-UP and UPF) within relevant slices and data center pools and sends reconfiguration requests through the O1 interface to adjust CPU and RAM allocations. Additionally, the RM-rApp issues A1 policies to the NS-xApp to enforce slice assurance.

\subsubsection{Network Slicing xApp (NS-xApp)}
The NS-xApp, residing in the Near-RT RIC, monitors slice-specific RAN performance, interprets and executes policies from the Non-RT RIC, and performs E2 actions like adjusting PRB allocation to meet RAN slice subnet requirements.

\section{Testbed Setup}
\label{sec:ExperimentSetup}

In our INA-Infra testbed, we use 3 VMs with different hardware capabilities to deploy as Edge, Regional, and Central DC pools with the simulated pay-as-you-go cost model of renting resources (CPU, RAM, network bandwidth), as defined in Tab.~\ref{tab:vm_resources_and_costs}. The low CPU allocation of Edge DC (500ms) is due to our simulated research environment with a small number of slices and UEs.

\begin{table}[h!]
\centering
\begin{tabular}{c|c|c|c}
\toprule
\textbf{Resource} & \textbf{Edge DC} & \textbf{Regional DC} & \textbf{Central DC} \\ 
\midrule
\textbf{CPU[ms]} & 500 & 2000 & 10000 \\ 
\midrule
\textbf{RAM (GB)} & 5 & 20 & 100 \\ 
\midrule
\textbf{CPU Cost (\$/100ms/h)} & 0.5 & 0.05 & 0.001 \\ 
\midrule
\textbf{RAM Cost (\$/GB/h)} & 0.1 & 0.01 & 0.002 \\ 
\midrule
\textbf{B/W Cost (\$/GB)} & 0.1 & 0.1 & 0.1 \\ 
\bottomrule
\end{tabular}
\caption{Hardware capabilities and resource costs.}
\label{tab:vm_resources_and_costs}
\vspace{-5mm}
\end{table}

\begin{table}[ht!]
\centering
\begin{tabular}{c|c|c | c}
\toprule
\textbf{Resources} & \textbf{CPU[ms]} & \textbf{RAM[MB]} & \textbf{TP[Mbps]} \\ \midrule
Nephio Workload Cluster & 150 & 1180 & NA \\ \midrule
\multirow{3}{*}{CU-UP deployment} & 0 & 2.8 & 0 \\ 
                         & 21 &  3.8 & 20 \\ 
                         & 244 &  3.8 & 250 \\ \midrule            
\multirow{3}{*}{UPF deployment} & 0 & 4.7 & 0 \\ 
                         & 27 &  4.8 & 20 \\ 
                         & 307 &  4.8 & 250 \\
\bottomrule
\end{tabular}
\caption{Resources usage under different traffic conditions.}
\label{tab:benchmark_resources}
\vspace{-2mm}
\end{table}


Each VM has a Nephio Workload Cluster to manage the NFs (Fig.~\ref{fig:nephio-archi}): Central DC for OAI 5G CN exclusively while all DCs for OAI RAN.  All network slices share the control plane deployment (i.e., CU-CP and AMF); but each slice will use separate data plane deployment (i.e., CU-UP, and UPF). Nephio helps deploy NFs in each slice via the O2 interface. Using USRP X410 as the RU, we ran experiments over-the-air (OTA) with 40 MHz bandwidth in the band N78 (center frequency 3425.01\,MHz) which we were legally allowed to use on our campus for research purposes.  The Pegatron 5G Dongles 
are used as UE for each network slice in the experiments below. Delays between DC pools are simulated using \texttt{traffic control~(tc)} Linux utility inside \texttt{containerlab} switches.

We benchmark the CPU and RAM usage of Nephio Workload Cluster (initial environment), as well as the CU-UP and UPF deployment during no traffic, low traffic (20 Mbps), and high traffic (250 Mbps) using \texttt{iperf3} as shown in Tab. \ref{tab:benchmark_resources}.

With INA-infra, MNOs define the high-level intention SLA (delay and throughput) in the NSMF, which will be translated to Nephio API to be realized in the O-Cloud platform. Additionally, our RM rApp algorithm (not included in this paper) will help manage RAN resources for NFs deployment while our NS-xApp will update the PRBs of DU via the E2 interface accordingly for every new slice.

\section{Experimental Results}
\subsection{Experiment 1 - Deployment Scenario}
\label{subsec:Experiment1}

In this scenario, we consider 4 network slices with different SLAs defined in Tab.~\ref{tab:SLA_Exp1}.
Slice S1, S2, and S3 have the same throughput requirements. But their delay requirements are increasingly from 50ms to 30ms to 10ms.  Slices S3 and S4 have the same delay requirement, but the S4 requires much more throughput than all other slices.

To meet these SLAs, the E2E delay is the main consideration when deciding where to deploy the NFs (CU-UP and UPF) of each slice. With the secondary goal of minimizing renting costs, S1's NFs can be deployed at the Central DC while S2's NFs can be deployed at the Regional DC while ensuring delay requirement. S3 and S4 NFs must be deployed at Edge DC to meet strict delay requirements. The cost of deployment in Tab. \ref{tab:SLA_Exp1} is calculated using values in Tab. \ref{tab:vm_resources_and_costs} and Tab. \ref{tab:benchmark_resources}. We could observe a clear trade-off between ensuring delay requirement SLA and renting cost across each DC region, with Edge DC being the most expensive but with the lowest delay guaranteed.

\begin{table}[ht!]
\centering
\begin{tabular}{c|c|c|c|c|c}
\toprule
& \multicolumn{2}{c|}{\textbf{High-level intention}} & \multicolumn{2}{c|}{\textbf{Low-level action}} & \\
\cline{2-5}
\addlinespace[0.25em]
\textbf{Slice} & \textbf{Delay[ms]} & \textbf{TP[Mbps]} & \textbf{DC} & \textbf{PRBs} & \textbf{Daily} \\ 
\textbf{(SST, SD)} & \textbf{(min/max)} & \textbf{(min/max)} & \textbf{Chosen} & \textbf{Alloc} & \textbf{Cost[\$]} \\
\midrule
S1 (1, 1) & 50 / 100 & 70 / 250 & Central & 30 & 8\\ \midrule
S2 (1, 2) & 30 / 70 & 70 / 250 & Regional & 30 & 10\\ \midrule
S3 (1, 3) & 10 / 50 & 30 / 250 & Edge & 15 & 22\\ \midrule
S4 (1, 4) & 10 / 50 & 45 / 250 & Edge & 20 & 28\\
\bottomrule
\end{tabular}
\caption{High-level intention and Low-level action for 4 slices.}
\label{tab:SLA_Exp1}
\vspace{-2mm}
\end{table}

\begin{figure}[!t]
\centering
\includegraphics[height=2in,trim=0 0 0 0.1in]
{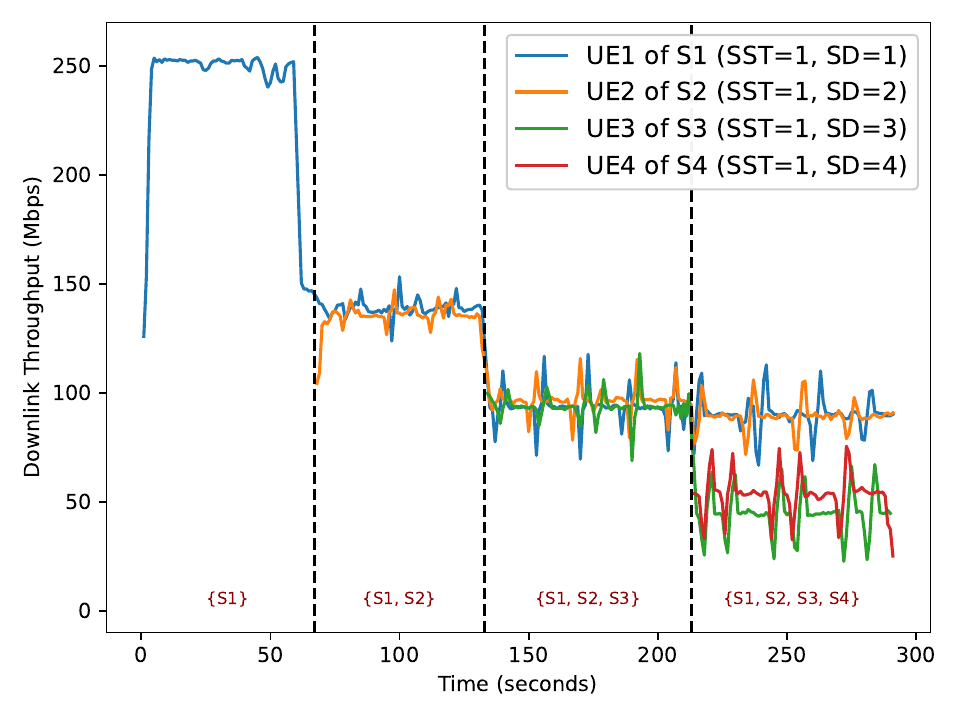} 
\caption{Throughput of 4 UEs in 4 network slices.}
\label{fig:Exp1_throughput}
\vspace{-2mm}
\end{figure}

In the experiment, we first start UE1 (i.e., Pegatron 5G dongle) associated with S1 (Central) and measure throughput via \texttt{iperf3}; then we slowly start other UEs associated with S2, S3, and S4 subsequently. Fig.~\ref{fig:Exp1_throughput} shows the average measured downlink (DL) throughput of the four UEs under the 4 four evaluated network slices. We can see that, first, on its own, the UE1 can achieve maximum throughput of the cell of 250 Mbps. When S2 (Regional) is started with UE2, both UEs share the total bandwidth, each can achieve 125 Mbps DL throughput. At 150s, S3 (Edge) is started with UE3, the NS-xApp will update the radio resource to support 3 slices, resulting in dropping the DL throughput of the three UEs to around 80 Mbps. Finally, when S4 (Edge) is introduced, the UE4 of S4 and UE3 are allocated around 50 Mpbs DL throughput each while UE1 and UE2 maintain 80 Mbps. During the whole experiment, all UEs achieve the throughput SLA defined in Tab.~\ref{tab:SLA_Exp1}.

\begin{figure}[!t]
\centering
\includegraphics[height=2in,trim=0 0 0 0.3in]
{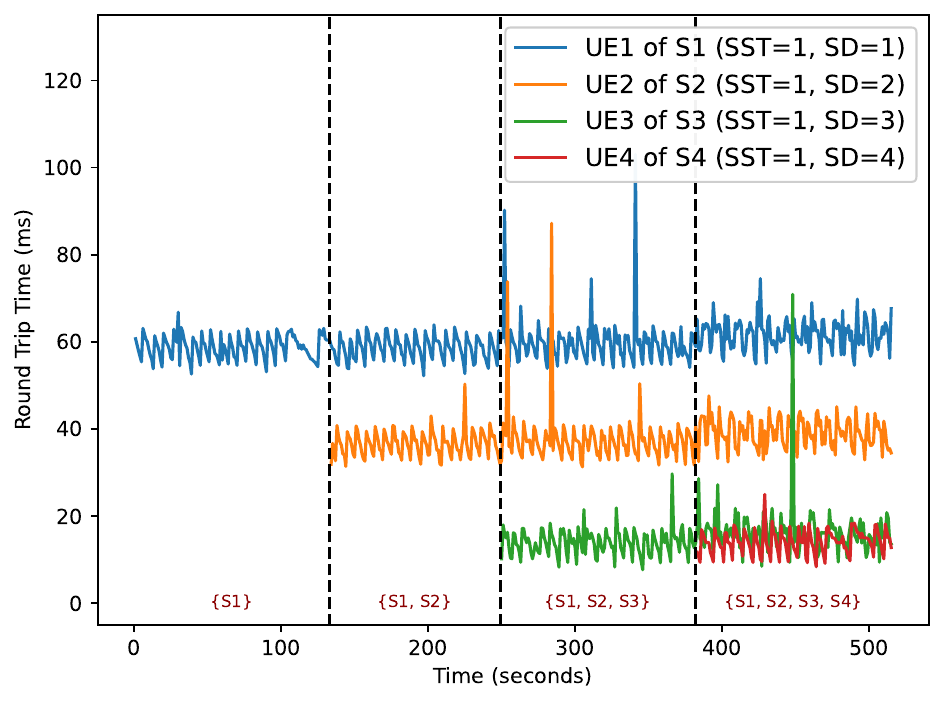} 
\caption{Round Trip Time of 4 UEs in 4 network slices.}
\label{fig:Exp1_RTT}
\vspace{-5mm}
\end{figure}

Fig.~\ref{fig:Exp1_RTT} shows round-trip-time (RTT) measured (via \texttt{ping} to N6's IP address) of the four UEs under 4 evaluated network slices. Note that we measure RTT under low traffic load conditions, i.e., not jointly test with \texttt{iperf3} throughput measurement, to ensure the latency is not affected by the high traffic load. As observed, RTT of the four UEs are met the required delay SLA defined in Tab.~\ref{tab:SLA_Exp1}.

\subsection{Experiment 2 - Configuration Scenario}
\label{subsec:Experiment2}

In the second scenario, we consider multiple slices deployed at a single DC (Edge) and overload its compute resources. As a result, without the management of compute resources, the performance of each slice NFs (CU-UP and UPF) deteriorates, causing a violation of the guaranteed SLA. The goal is to minimize SLA violations with RM rApp.

\begin{figure}[H]
\centering
\begin{subfigure}[b]{\columnwidth}
    \centering
    \includegraphics[height=2in]{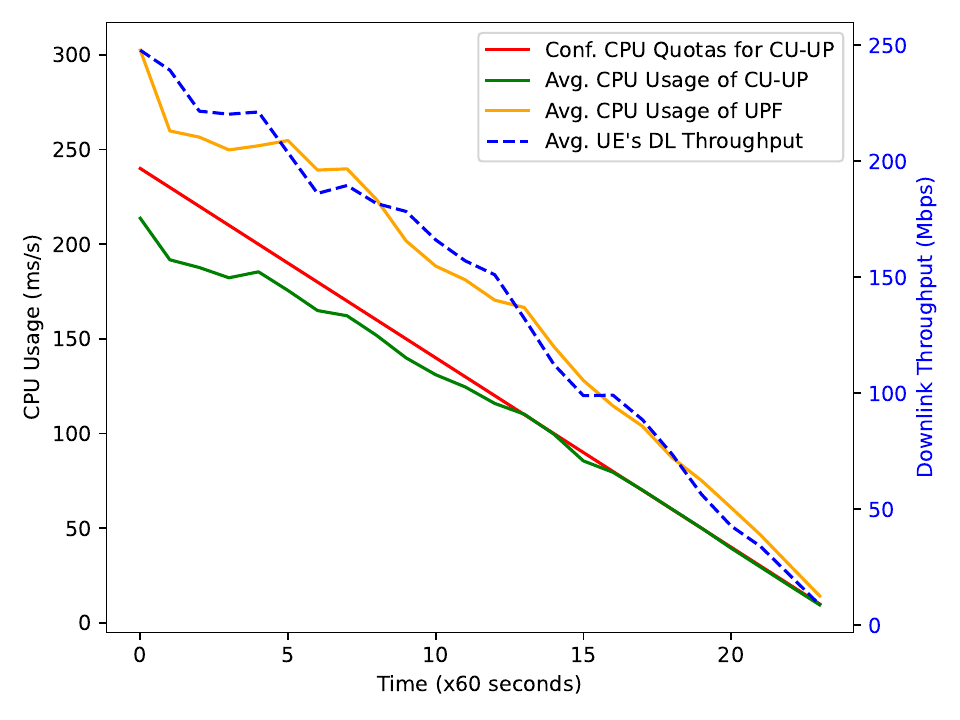}
    \caption{CU-UP's CPU quotas effect on CU-UP, UPF and UE's DL throughput}
    \label{fig:CU-UP_CPU_Usage_vs_Throughput-a}
\end{subfigure}
\begin{subfigure}[b]{\columnwidth}
    \centering
    \includegraphics[height=2in]{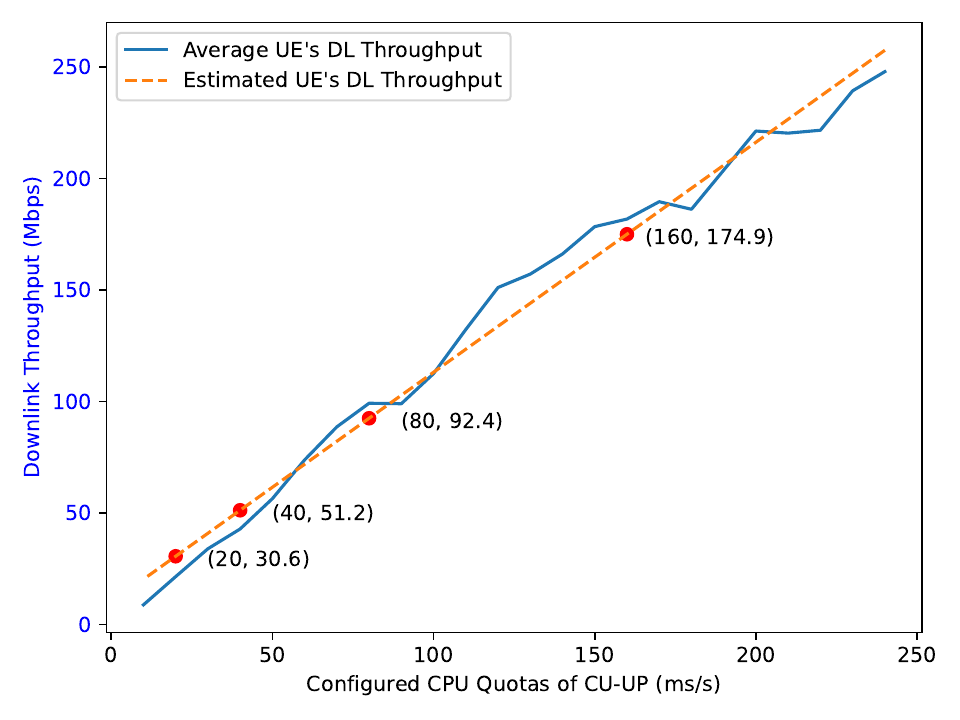}
    \caption{UE's DL throughput versus CU-UP's CPU quotas}
    \label{fig:CU-UP_CPU_Usage_vs_Throughput-b}
\end{subfigure}
\caption{Investigate CPU quotas effect on UE's DL throughput}
\vspace{-2mm}
\label{fig:CU-UP_CPU_Usage_vs_Throughput}
\end{figure}

We first investigate how the compute resources affect UE's DL throughput by keeping the radio resource constant while lowering the CPU quotas (ms) of the CU-UP over time (Fig. \ref{fig:CU-UP_CPU_Usage_vs_Throughput-a}). As observed, the CPU usage of both CU-UP and UPF would decrease with the lower CU-UP's CPU quotas, which in turn lower the UE's DL throughput. This linear relationship is further confirmed when plotting UE's DL throughput against the CU-UP configured CPU quotas (Fig. \ref{fig:CU-UP_CPU_Usage_vs_Throughput-b}).

After the initial investigation, we deploy an additional slice S5 with the highest priority and other SLA metrics as defined in Tab.~\ref{tab:SLA_Exp2}. Similar to Experiment 1, S5 would be deployed in the Edge DC due to its low delay SLA (i.e., 10ms).
Note that among three slides deployed at the Edge DC (i.e., S3, S4, and S5), S5 has higher priority. 
Compared to Experiment 1, the throughput (and projected into PRBs) of slice S5 is reallocated from those of S1 and S2 slices. 

\begin{table}[h]
\centering
\begin{tabular}{c|c|c|c|c|c}
\toprule
& \multicolumn{2}{c|}{\textbf{High-level intention}} & \multicolumn{2}{c|}{\textbf{Low-level action}} & \\
\cline{2-5}
\addlinespace[0.25em]
\textbf{Slice} & \textbf{Delay[ms]} & \textbf{TP[Mbps]} & \textbf{DC} & \textbf{PRBs} & \textbf{Priority} \\ 
\textbf{(SST, SD)} & \textbf{(min/max)} & \textbf{(min/max)} & \textbf{Chosen} & \textbf{Alloc} & \\
\midrule
S1 (1, 1) & 50 / 100 & 25 / 250 & Central & 10 & 1\\ \midrule
S2 (1, 2) & 30 / 70 & 25 / 250 & Regional & 10 & 1\\ \midrule
S3 (1, 3) & 10 / 50 & 30 / 250 & Edge & 15 & 2\\ \midrule
S4 (1, 4) & 10 / 50 & 45 / 250 & Edge & 20 & 2\\ \midrule
S5 (1, 5) & 10 / 50 & 90 / 250 & Edge & 40 & 3\\
\bottomrule
\end{tabular}
\caption{High-level intention and Low-level action for 5 slices. \\A higher priority value is more important.}
\label{tab:SLA_Exp2}
\vspace{-2mm}
\end{table}

\begin{figure*}[t]
\centering
\begin{subfigure}[t]{0.32\textwidth}
    \centering
    \includegraphics[width=\textwidth]{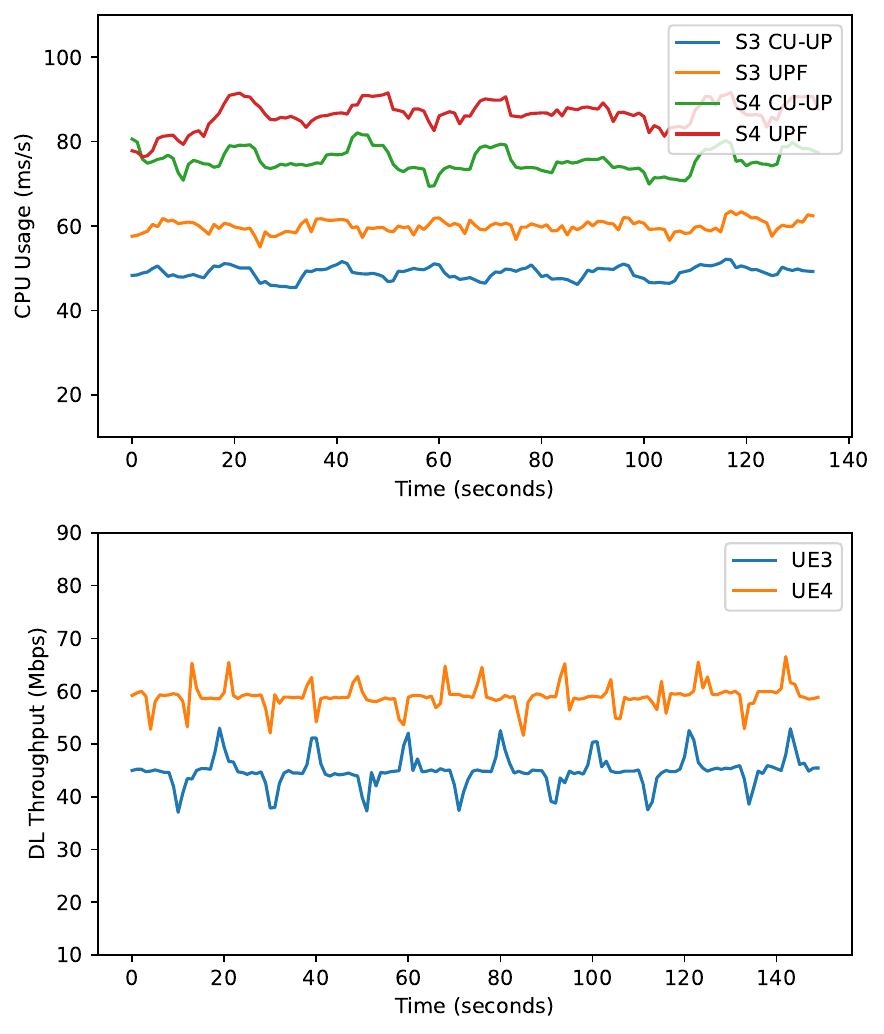}
    \caption{Without S5.}
    \label{fig:Exp2_Result_withoutS5}
\end{subfigure}%
\hfill
\begin{subfigure}[t]{0.32\textwidth}
    \centering
    \includegraphics[width=\textwidth]{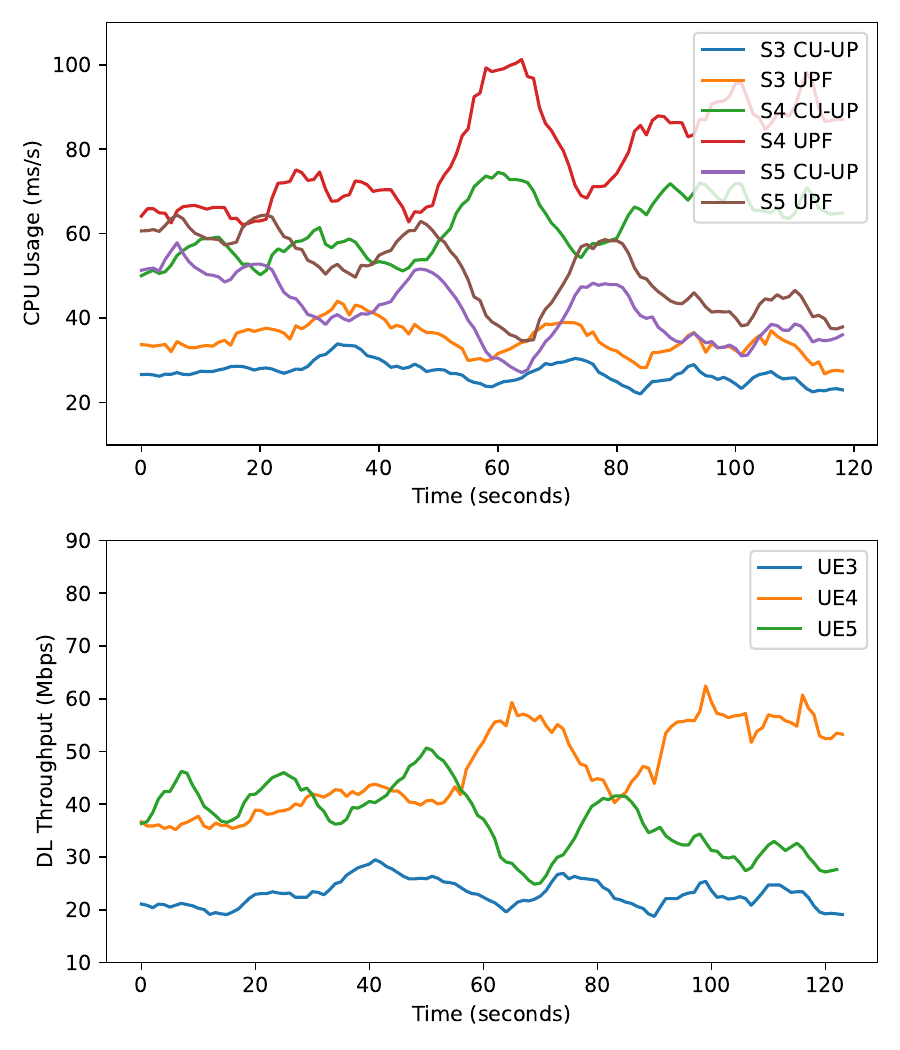}
    \caption{With S5 but without RM rApp.}
    \label{fig:Exp2_Result_withS5_without_RM}
\end{subfigure}%
\hfill
\begin{subfigure}[t]{0.32\textwidth}
    \centering
    \includegraphics[width=\textwidth]{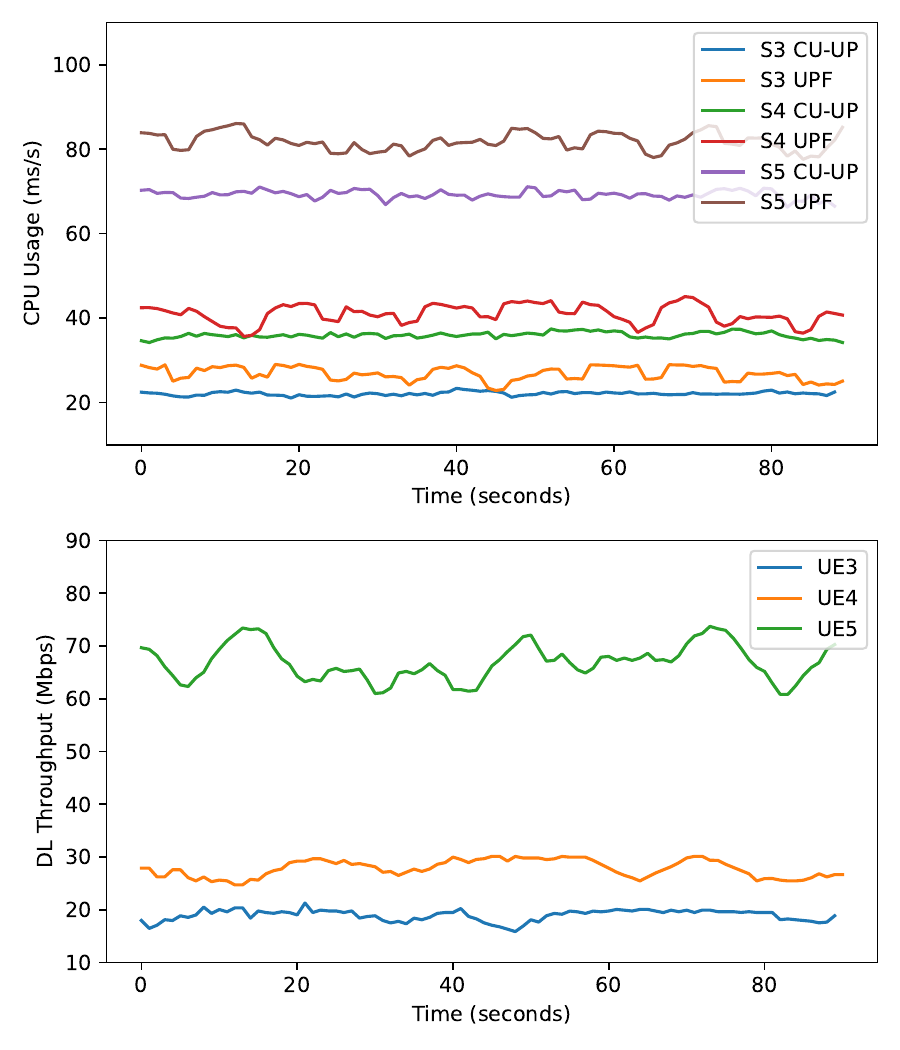}
    \caption{With S5 and with RM rApp.}
    \label{fig:Exp2_Result_withS5_with_RM}
\end{subfigure}
\caption{Comparison of Edge Cluster CPU Usage and Throughput under different scenarios.}
\label{fig:Exp2_Result}
\vspace{-5mm}

\end{figure*}

Fig.~\ref{fig:Exp2_Result} shows average DL throughput (bottom) and CPU usage of CU-UP and UPF deployments (top) of the slices at the Edge DC. 
Initially, Edge DC only had 2 slices (Fig.~\ref{fig:Exp2_Result_withoutS5}), S3 and S4 both satisfying its SLA requirement with average DL throughput of 45 and 60 Mbps, respectively. Next, with the introduction of S5 (Fig.~\ref{fig:Exp2_Result_withS5_without_RM}), when there's no resource management from rApp, we could observe the throughput of all three slices drop due to lack of compute resources on CU-UP and UPF. In this case, S5 with the highest priority faces severe SLA violation (a 67\% drop from 90 to 30 Mbps). Finally, using RM rApp (Fig.~\ref{fig:Exp2_Result_withS5_with_RM}), even though all 3 slices S3, S4, S5 have SLA violation with average DL throughput of 15, 25 and 65 Mbps, respectively; S5 being the highest priority (3) has the lowest SLA violation (28\% compared to 50\% and 45\% for S3 and S4).


\section{Conclusion}
This paper develops INA-Infra, an open, extensible, and end-to-end 5G/beyond 5G network infrastructure with intent-driven network automation and end-to-end network slicing capability. Our experiments showcase the realization of mobile network operators' high-level intentions into low-level actions and deployments, using Nephio infrastructure and network resources automation as well as intelligent control with Resource Management rApp and Network Slicing xApp. 

\section*{Acknowledgment}
This research is supported by the National Research Foundation, Singapore and Infocomm Media Development Authority under its Future Communications Research \& Development Programme. Any opinions, findings and conclusions or recommendations expressed in this material are those of the author(s) and do not reflect the views of National Research Foundation and Infocomm Media Development Authority, Singapore.

\bibliographystyle{IEEEtran}
\bibliography{ref}
\end{document}